\documentclass[12pt,onecolumn]{article}
\usepackage{amsmath}
\usepackage{amssymb}
\usepackage{graphicx}
\usepackage{caption2}
\usepackage{amsfonts}
\usepackage{cite}

\newcommand{\be}{\begin{equation}}
\newcommand{\ee}{\end{equation}}
\newcommand{\bea}{\begin{eqnarray}}
\newcommand{\eea}{\end{eqnarray}}
\newcommand{\bef}{\begin{figure}}
\newcommand{\ef}{\end{figure}}
\newcommand{\bt}{\begin{tabular}}
\newcommand{\et}{\end{tabular}}
\newcommand{\bno}{\begin{enumerate}}
\newcommand{\eno}{\end{enumerate}}

\def\3{\ss}

\catcode`\"=\active
\def"{\accent'177}

\begin{document}

\begin{center}
{\bf\large On cusped solitary waves in finite water depth}

Shijun Liao

State Key Laboratory of Ocean Engineering 

School of Naval Architecture,Ocean and Civil Engineering\\  Shanghai Jiaotong University, Shanghai 200240, China

Department of Mathematics, Shanghai Jiaotong University, Shanghai 200240, China  

( Email address: sjliao@sjtu.edu.cn )

\end{center}

\begin{abstract}
It is well-known that the Camassa-Holm (CH) equation admits both of the peaked and cusped solitary waves in shallow water.  However, it was an open question whether or not the exact wave equations can admit them in finite water depth.  Besides, it was traditionally believed that  cusped solitary waves, whose 1st-derivative tends to infinity at crest,  are essentially different from peaked solitary ones with finite 1st-derivative.   Currently, based on the symmetry and the exact water wave equations, Liao \cite{Liao-UWM} proposed a unified wave model (UWM) for progressive gravity waves in finite water depth.  The UWM admits not only all traditional smooth progressive waves but also the peaked solitary waves in finite water depth: in other words,   the peaked solitary progressive waves are consistent with the traditional smooth ones.   In this paper,  in the frame of the linearized UWM, we further give, for the first time,  the cusped solitary waves in finite water depth,  and  besides  reveal a close relationship between the  cusped and peaked solitary waves:  a cusped solitary wave is consist of an infinite number of peaked solitary ones with the same phase speed, so that it can be regarded as a special peaked solitary wave.   This also well explains  why and how a cuspon has an infinite 1st-derivative at crest.   It is found that, like peaked solitary waves,   the vertical velocity of a cusped solitary wave in finite water depth is also discontinuous at crest ($x=0$), and  especially  its  phase  speed  has nothing to do with wave height, too.  In addition, it is unnecessary to consider whether  the peaked/cusped solitary waves given by the UWM are weak solution or not, since the governing  equation  is not necessary to  be  satisfied  at crest.    All of these would deepen and enrich our understandings about the cusped solitary waves.
\end{abstract}



\setlength{\parindent}{0.75cm} 

\section{\label{sec:introduction}Introduction }

 The  smooth solitary surface wave  was  first reported by John Scott Russell \cite{Russell1844} in 1844.  Since then,    various types of solitary waves have been found.   The mainstream models of shallow water waves, such as the Boussinesq equation  \cite{Boussinesq1872}, the KdV equation  \cite{KdV}, the  BBM equation  \cite{Benjamin1972} and so on, admit  dispersive  {\em smooth}  periodic/solitary  progressive waves with permanent form: the wave elevation is {\em infinitely} differentiable {\em everywhere}.   Especially,  the phase  speed  of  the smooth waves is  highly dependent upon wave height:   the larger the wave height of a  smooth progressive wave,  the faster it propagates.  Nowadays,  the  smooth amplitude-dispersive  periodic/solitary waves  are the mainstream  of  researches  in  water waves.

In 1993, Camassa and Holm \cite{Camassa1993PRL} proposed the celebrated Camassa-Holm (CH) equation for shallow water waves, and first reported  the  so-called peaked solitary wave, called peakon,  which has a peaked crest with a discontinuous (but finite) 1st-order derivative at crest.  This is a breakthrough in water wave theories, since it opens a new field of research in the past 20 years.    Physically, different from the KdV equation and Boussinesq equation,  the  CH equation   can model  phenomena of not only soliton interaction but also wave breaking  \cite{Constantin2000}.   Mathematically,  the CH equation is integrable and bi-Hamiltonian,  therefore  possesses an infinite number of conservation laws in involution  \cite{Camassa1993PRL}.   Besides,  it is associated with the geodesic flow on the infinite dimensional Hilbert manifold of diffeomorphisms of line  \cite{Constantin2000}.    Thus, the CH equation  has lots of  intriguing  physical  and  mathematical properties.   It is even believed that  the CH equation  ``has the potential to become the new master equation for shallow water wave theory'' \cite{Fushssteiner1996}.    In addition,  Kraenkel and Zenchuk  \cite {Kraenkel1999}  reported  the  cusped  solitary waves  of the CH equation, called cuspon.    The so-called  cuspon is a kind of solitary wave with the 1st derivative  going  to  {\em infinity} at crest.   Note that, unlike a peakon that has a {\em finite} 1st derivative, a cuspon has an {\em infinite} 1st derivative at crest.  Thus, it was traditionally believed that peakons and cuspons are completely different two kinds of solitary waves.  

However, the CH equation is a simplified model of water waves in shallow water.   It was an open question whether or not the exact wave equations admit the peaked and cusped solitary waves in {\em finite} water depth.   For example,  the velocity distribution of peaked/cusped solitary waves in the vertical direction was  unknown, since it can not be determined by a wave model in shallow water (such as the CH equation).   Currently,  based on the symmetry and the exact wave equations,  Liao  proposed a unified wave model (UWM)  for progressive gravity waves in finite water depth with permanent form \cite{Liao-UWM}.   It was found that the UWM admits not only all traditional smooth periodic/solitary waves but also the peaked solitary waves in finite water depth, even including the famous peaked solitary waves of the CH equation as its special case.   Therefore, the UWM unifies both of the smooth and peaked solitary waves in finite water depth, for the first time.  In other words, the progressive peaked solitary waves in finite water depth  are consistent with the traditional smooth waves, and thus are as acceptable and reasonable  as the smooth ones.  

In this article, using the linearized UWM, we give an closed-form expression of cusped solitary waves in finite water depth, and illustrate that a cusped solitary wave is consist of an infinite number of peaked solitary ones.   This reveals, for the first time to the best of my knowledge,  a simple but elegant relationship between the peaked and cusped solitary waves in finite water depth.    

\section{Cusped solitary waves in  finite water depth}

Let us first describe the UWM  briefly.   Consider a  progressive gravity wave propagating on a horizontal  bottom  in a {\em finite} water depth $D$, with a constant phase speed $c$ and a  permanent  form.   For simplicity,  the  problem  is solved  in the frame moving with the phase speed $c$.   
Let $x, z$ denote the horizontal  and  vertical dimensionless co-ordinates (using the water depth $D$ as the characteristic length),   with  $x=0$ corresponding to the wave crest,  $z=-1$ to the bottom, and  the $z$  axis upward, respectively.      Assume that the wave elevation $\eta(x)$ has a symmetry about the crest,  the fluid  in  the  interval  $x >0$ is inviscid and incompressible,  the flow   in $x >0$  is irrotational, and surface tension is neglected.   Here, it should be emphasized that, different from all traditional wave models,   the flow at $x=0$ is {\em not}  absolutely  necessary  to be irrotational.   Let   $\phi(x,z)$ denote the velocity potential.    All of them are dimensionless using $D$  and  $\sqrt{g D}$ as the characteristic  scales of length  and  velocity, where $g$ is the acceleration due to   gravity.   In the frame of the UWM, the velocity potential $\phi(x,z)$ and the wave elevation $\eta(x)$ are first determined by the exact wave equations (i.e. the Laplace equation $\nabla^2 \phi = 0$, the two nonlinear boundary conditions on the unknown free surface $\eta$, the bed condition and so on) only in the interval  $x\in(0,+\infty)$, and then extended to the whole interval $(-\infty, +\infty)$ by means of the symmetry
 \[
 \eta(-x) = \eta(x),  u(-x,z) = u(x,z),  v(-x,z) = -v(x,z),
 \] 
 which enforces the additional  restriction condition $v(0,z)=0$.  Note that, in the frame of the UWM,  the flow at $x = 0$ is {\em not} necessarily irrotational,  so that  the UWM is more general: this is the reason why the UWM can admit both of the smooth and peaked solitary waves.    
 
In the interval $(0, +\infty)$, the governing equation $\nabla^2 \phi(x,z) = 0$ with the bed condition $\phi_z(x,-1)=0$ has two kinds of general solutions \cite{Mei-book}, where the subscript denotes the differentiation with respect to $z$.  One is  \[  \cosh[n k (1+z)]\sin(n kx),\]  corresponding to the smooth periodic waves with the dispersive relation 
\begin{equation}
\alpha^2 = \frac{\tanh(k)}{k} \leq 1,  \label{geq:k:smooth}
\end{equation} 
where $\alpha = c/\sqrt{g D}$ is the dimensionless phase speed,  $k$ is wave number and $n$ is an integer, respectively.  The other is \[  \cos[n k (z+1)] \exp(- n k x),\] corresponding to the peaked solitary waves in finite water depth \cite{Liao-UWM}, with the relation
 \begin{equation}
 \alpha^2 = \frac{\tan(k)}{k} \geq 1,  \label{geq:k:peaked}
 \end{equation} 
 where $k$ has nothing to do with wave number.  
 Given $\alpha\leq 1$  for the smooth periodic waves, the transcendental equation (\ref{geq:k:smooth}) has a {\em unique} solution, as mentioned in the textbook  \cite{Mei-book}.   However, given $\alpha\geq 1$ for the peaked solitary waves, the  transcendental equation (\ref{geq:k:peaked}) has an {\em infinite} number of solutions:         
\begin{equation}
\alpha^2 =\frac{\tan k_n}{k_n}, \hspace{0.5cm}   n \pi \leq k_{n} \leq n \pi +\frac{\pi}{2}, \;\; n\geq 0, \label{geq:k}
\end{equation}  
corresponding to an infinite number of peaked solitary waves  \cite{Liao-UWM}
\begin{equation}
\eta_n(x) = A_n \; \exp(-k_n \; |x|)   \label{eta:peaked}
\end{equation}
in the frame of the linear UWM,  where $A_n$ denotes its wave height.   For example, when $\alpha^2 = 2\sqrt{3}/\pi$,  the  transcendental equation (\ref{geq:k:peaked}) has an infinite number of solutions  $k_0=\pi/6$,   $k_1 = 4.51413$, $k_2 = 7.73730$, $k_3 = 10.91266$, $k_4 = 14.07281$, $k_5 = 17.22616$, $k_6 = 20.37587$, $k_7 = 23.52341$, $k_8=26.66955$, $k_9 = 29.81472$, $k_{10} = 32.95921$, and the asymptotic expression  
\begin{equation}
k_n \approx \left( n +0.5 \right)\pi, \hspace{0.5cm} n >10,     \label{k:approximate}
\end{equation} 
 with less than 0.08\% error.  In general, $k_n \approx (n+0.5)\pi$ is a rather accurate approximation of $k_n$ for large enough integer  $n$.  
 
Obviously, the peaked solitary wave (\ref{eta:peaked}) is not smooth at crest, i.e. its first derivative is discontinuous.  Note that the well-known peaked solitary wave $\eta = c \; \exp (-|x|)$ of the CH equation is only a special case of (\ref{eta:peaked}) when $A_n = c$ and $k_n=1$.    However,  unlike  $\eta = c\; \exp(-|x|)$ that is a weak solution of the CH equation,   it is {\em unnecessary} to consider whether or not the peaked solitary wave (\ref{eta:peaked})  is a kind of  weak solution, because the CH equation is defined in the whole domain $-\infty < x < +\infty$  but the governing equation of the UWM is defined only in $0<x < +\infty$.   Physically,  unlike the CH equation and the fully nonlinear wave equations, waves in the frame of the UWM are {\em not} necessary to be irrotational at $x=0$, therefore the governing equation  holds  only in  the  domain $0<x<+\infty$, since the solution in the interval $-\infty < x <0$ is gained by means of the symmetry.   Mathematically, $x=0$ is a boundary of the governing equation, and it is well-known that solutions  of  differential  equations  can be non-smooth at boundary, like a  beam with discontinuous  cross  sections acted by a constant bending moment.  Therefore, in the frame of the UWM, it is {\em unnecessary}  to  consider  whether  or not  the peaked solitary waves (\ref{eta:peaked}) are weak solutions at all.   This is the reason why, unlike the well-known peaked solitary wave $\eta = c\; \exp(-|x|)$ of the CH equation whose phase speed is {\em always} equal to its wave height,  the phase speed of the peaked solitary waves   (\ref{eta:peaked})  given by the UWM has {\em nothing} to do with wave height!  This  is the most attractive   novelty of the UWM.    

The above peaked solitary waves in finite water depth have some unusual characteristics, as revealed by Liao  \cite{Liao-UWM}.  First, it has a peaked crest with a discontinuous vertical velocity $v$ at crest.  Besides,  unlike the smooth waves whose horizontal velocity $u$ decays exponentially from free surface to bottom,  the horizontal velocity $u$ of the peaked solitary waves at bottom is always larger than that on free surface.  Especially,  different from the smooth waves whose phase speed depends upon wave height, the phase speed of the peaked solitary waves in finite water depth has nothing to do with wave height, i.e. it is {\em non-dispersive}.            

 \begin{figure}
\centering
\includegraphics[scale=0.55]{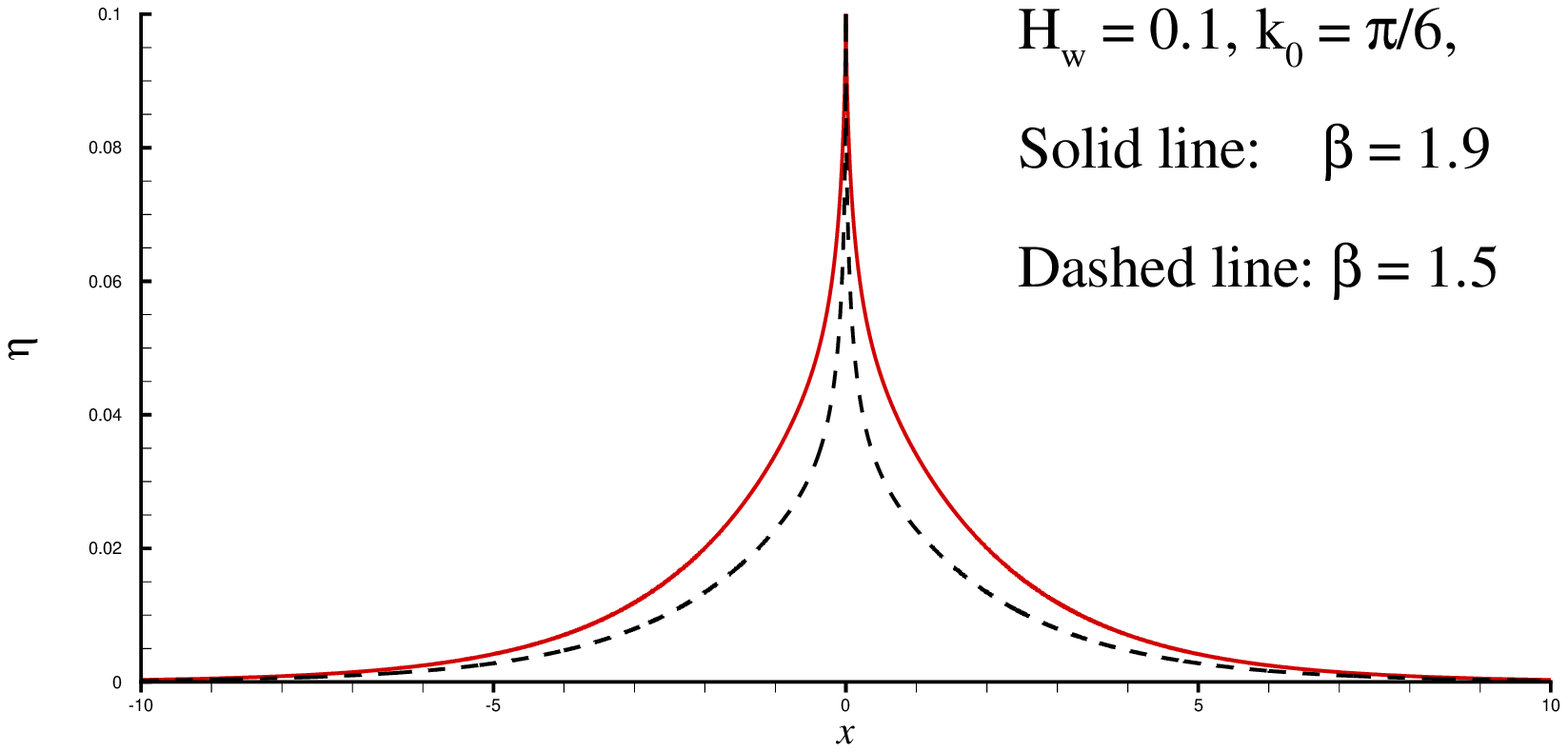}
\caption{Cusped solitary waves in finite water depth defined by (\ref{def:cuspon:A}) when $H_w=0.1$ and $k_0=\pi/6$ (corresponding to $\alpha =12^{1/4}/\sqrt{\pi}$). Solid line: $\beta=1.9$; Dashed line: $\beta=1.5$. }
\label{figure:cuspon}

\centering
\includegraphics[scale=0.55]{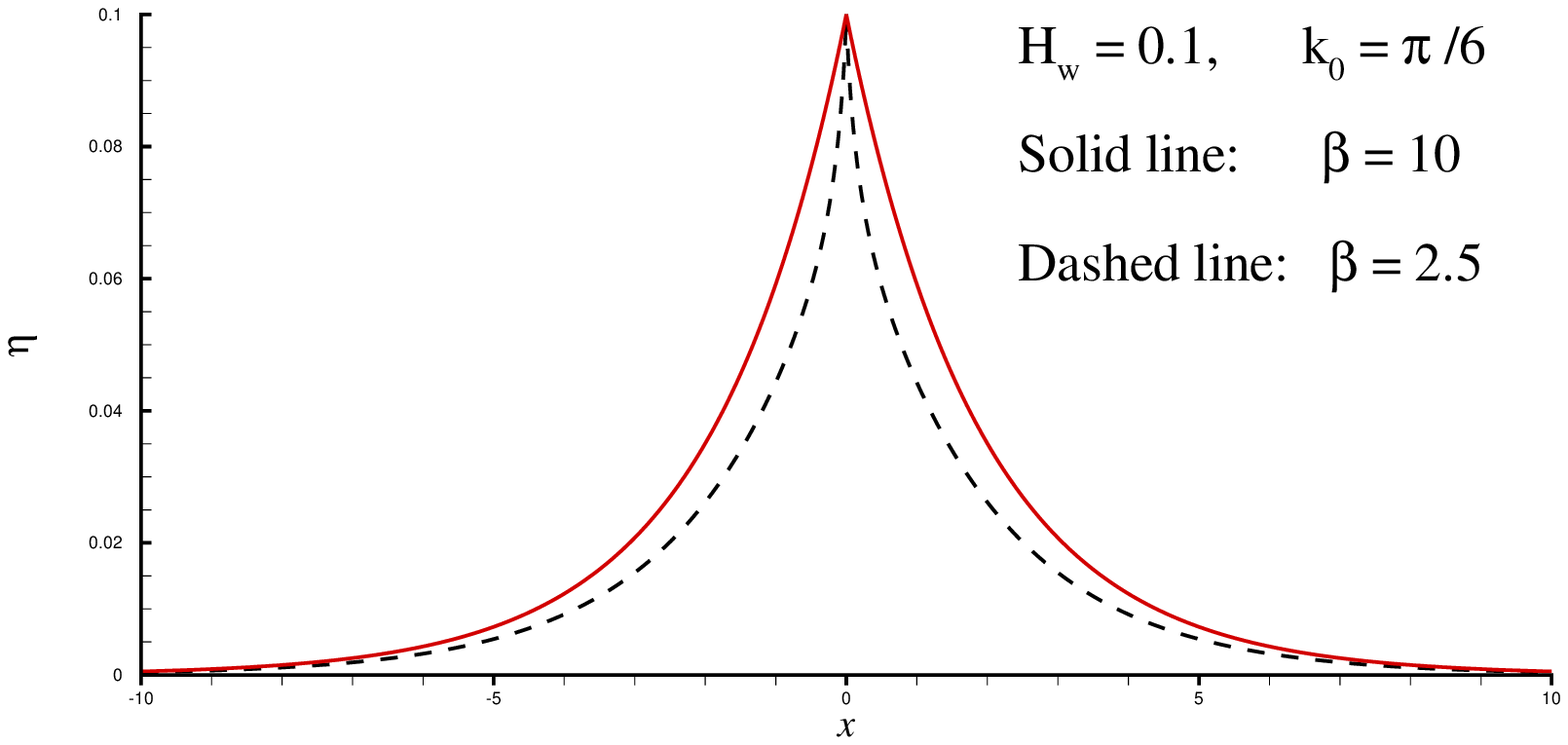}
\caption{Peaked solitary waves in finite water depth defined by (\ref{def:cuspon:A}) when $H_w=0.1$ and $k_0=\pi/6$ (corresponding to $\alpha =12^{1/4}/\sqrt{\pi}$). Solid line: $\beta=10$; Dashed line: $\beta=2.5$. }
\label{figure:peakon}
\end{figure}

Thus, in the frame of the linear UWM \cite{Liao-UWM},  given a dimensionless phase speed  $\alpha\geq 1$,  there exist an {\em infinite} number of peaked solitary waves $A_n \exp(-k_n |x|)$ with the {\em same} phase speed $\alpha$ but different wave amplitudes $A_n$.   Thus,  we may have such peaked solitary waves  
\[   \eta(x) = \sum_{n=0}^{\infty} A_n \; \exp(-k_n |x|),   \]
where $A_n$ is a constant, which can be chosen with great freedom, as long as the above infinite series is convergent in the whole interval $(-\infty, +\infty)$.     
As a special case of it, let us consider such a one-parameter family of wave elevations  
\begin{equation}
\eta(x) = \frac{H_w}{\zeta(\beta)}\sum_{n=1}^{+\infty} \frac{1}{n^\beta} \; \exp(-k_{n-1} |x|),  \;\; \beta >1, \label{def:cuspon:A}
\end{equation}   
where $H_w$ denotes wave height,  $\beta >1$ is a constant,  $\zeta(\beta)$ is the Riemann zeta function, and $k_n$ is determined by (\ref{geq:k:peaked}) for the given $\alpha \geq 1$,  respectively.   Since $\beta>1$, we have $ \sum_{n=1}^{+\infty} n^{-\beta}=\zeta(\beta)$ so that the above infinite series converges to the wave height $H_w$ at $x=0$, and besides is convergent in the whole interval $(-\infty, +\infty)$.  However, its 1st  derivative at $x=0$, i.e.       
\begin{equation}
\eta' (0) = \pm \frac{H_w}{\zeta(\beta)}\sum_{n=1}^{+\infty} \frac{k_{n-1}}{n^\beta},
\end{equation} 
is convergent to a {\em finite} value when $\beta>2$, but  tends to {\em infinity} when $1 <\beta  \leq  2$,  because  $k_{n-1} \approx (n-0.5)\pi$ for large enough integer $n$ and the series $\sum 1/n^{\beta-1}$ is convergent when $\beta>2$  but tends to infinity when $0< \beta \leq 2$.   So,  the infinite series (\ref{def:cuspon:A}) defines a {\em cusped} solitary wave in finite water depth when $1<\beta\leq 2$  and a {\em peaked} solitary wave when $\beta>2$.    Therefore,  in essence,  a cusped solitary wave in finite water depth is consist of an {\em infinite} number of peaked solitary waves (when $1<\beta\leq 2$) with the same phase speed!    To the best of the author's knowledge, this reveals,  for the first time,  a simple but elegant relationship between the peaked and cusped solitary waves in finite water depth!  In addition, the infinite series (\ref{def:cuspon:A}) illustrates the consistency of the peaked and cusped solitary waves, and besides  explains  {\em why} and {\em how} a cuspon has an infinite 1st-derivative at crest.   Since the phase speed of peaked solitary waves (\ref{eta:peaked})  in finite water depth has nothing to do with the wave height,  it is straight forward that the phase speed of a cusped solitary wave in finite water depth also has nothing to do with the wave height, too.       

Note that,  according to the definition of the wave elevation (\ref{def:cuspon:A}),  given a dimensionless phase velocity $\alpha\geq 1$ and an arbitrary wave height $H_w$, there exist an {\em infinite} number of cusped solitary waves, dependent upon $\beta\in(1,2]$.   For example, the two cusped solitary waves in finite water depth defined by the infinite series (\ref{def:cuspon:A})  in the case of $\alpha^2 = 2\sqrt{3}/\pi, H_w = 1/10$ when $\beta=1.5$ and $\beta=1.9$ are as shown in Fig.~\ref{figure:cuspon}.  It should be emphasized that the {\em same} expression (\ref{def:cuspon:A})  can define an {\em infinite} number of peaked solitary waves in finite water depth, too, depending on $\beta\in(2,+\infty)$.   For example, the two peaked solitary waves in finite water depth in the case of $\alpha^2 = 2\sqrt{3}/\pi, H_w = 1/10$ when $\beta=5/2$ and $\beta=10$ are as shown in Fig.~\ref{figure:peakon}.     This well illustrates the consistency of the peaked and cusped solitary waves in finite water depth.     

Theoretically speaking, given an arbitrary wave height $H_w$ and a dimensionless phase speed $\alpha\geq 1$, there are many different types of peaked/cusped solitary waves in finite water depth.  For example,  a more generalized,  two-parameter family of  peaked/cusped  solitary waves in finite water depth reads   
\begin{equation}
\eta(x) =  \frac{H_w}{\zeta(\beta,\gamma)}\sum_{n=0,n\neq -\gamma}^{+\infty} \frac{1}{(n+\gamma)^\beta} \; \exp(-k_n |x|),   
\end{equation}
where $\beta>1$ and $\gamma \neq 0$ are constants to be chosen with great freedom,  $\zeta(\beta,\gamma)$ is a generalized Riemann zeta function, and $k_n$ is determined by (\ref{geq:k:peaked}) for the given $\alpha\geq 1$,  respectively.  Since $k_n \approx (n+0.5)\pi$ for large enough integer $n$,  the above infinite series defines a {\em cusped} solitary wave when $1<\beta\leq 2$ and a {\em peaked} ones when $\beta > 2$, respectively.   This illustrates once again the consistency of the peaked and cusped solitary waves in finite water depth.    

According to the linearized UWM \cite{Liao-UWM}, the velocity potential $\phi^+$ (defined only in the interval $x>0$) corresponding to  the peaked/cusped solitary wave elevation  (\ref{def:cuspon:A}) reads 
\begin{eqnarray}
\phi^+ &=& -\frac{\alpha H_w}{\zeta(\beta)} \sum_{n=1}^{+\infty}\frac{\cos[k_{n-1}(z+1)] \exp(-k_{n-1} x)}{n^\beta \; \sin(k_{n-1})},
\end{eqnarray}  
which gives, using the symmetry, the corresponding horizontal velocity 
 \begin{eqnarray}
 u &=& \frac{\alpha \; H_w}{\zeta(\beta)} \sum_{n=0}^{+\infty} \frac{k_{n} \cos[k_{n}(z+1)] \exp(-k_{n} |x|)}{(n+1)^\beta \; \sin(k_{n})} \;\;\; \;\;\;
 \end{eqnarray}
in the whole interval  $x\in(-\infty,+\infty)$,  the vertical velocity 
  \begin{eqnarray}
 v^+ &=& \frac{\alpha \; H_w}{\zeta(\beta)} \sum_{n=0}^{+\infty} \frac{k_{n} \sin[k_{n}(z+1)] \exp(-k_{n} x)}{(n+1)^\beta \; \sin(k_{n})}
 \end{eqnarray}
 in the interval $x\in(0,+\infty)$, and the vertical velocity 
   \begin{eqnarray}
 v^- &=& - \frac{\alpha \; H_w}{\zeta(\beta)} \sum_{n=0}^{+\infty} \frac{k_n \sin[k_{n}(z+1)] \exp( k_{n} x)}{(n+1)^\beta \; \sin(k_{n})} 
 \end{eqnarray}
in the interval $(-\infty,0)$, respectively.   Obviously, it holds \[  \lim_{x \to 0} v^+ = - \lim_{x \to 0} v^-\]  for  the  cusped ($1<\beta\leq 2$) and peaked ($\beta>2$) solitary waves,  although  we always have $v=0$ at $x=0$.   Thus, like a peakon in finite water depth,  a cuspon in finite water depth has the velocity discontinuity at $x=0$, too.   

Especially, at $z=0$ and as $x\to 0$,  the corresponding vertical velocity reads  
\[ 
\lim_{x\to 0}v^+(x,0)  =  \frac{\alpha \; H_w}{\zeta(\beta)} \sum_{n=0}^{+\infty} \frac{k_n}{(n+1)^\beta},
\]  
which is finite when $\beta>2$ but tends to infinity when $1<\beta \leq 2$,  since $k_n  \approx  (n+0.5)\pi$ for large enough integer  $n$.   Thus, unlike a peaked solitary wave in finite water depth whose  $v$ is always finite,  the vertical velocity of the cusped solitary waves in finite water depth tends to infinity at $z=0$ as $x\to 0$.   Mathematically,  this  is acceptable, since it is traditionally believed that a cuspon has a higher singularity than a peakon.   Such kind of singularity leads to a more strong vortex sheet at $x=0$ near $z=0$.   Physically,  in reality such kind of singularity and discontinuity,  ``if it could ever be originated, would be immediately abolished by viscosity'',  as mentioned by Lamb \cite{Lamb}.    

\section{Concluding remarks}

In summary, in the frame of the linearized UWM  \cite{Liao-UWM}, we give, for the first time,  the cusped solitary waves in finite water depth, and reveal that a cuspon is consist of an infinite number of peaked solitary waves with the same phase speed.  This kind of consistency also well explains why and how the 1st-derivative of a cusped solitary wave tends to infinity at crest.   It is found that, like a peakon,  the vertical velocity of a cuspon is also discontinuous  at $x=0$, and besides, its phase speed also has nothing to do with wave height, too.   All of these would deepen and enrich our understandings about the peaked and cusped solitary waves.      

It should be emphasized that, in the frame of the UWM,   the governing equation is defined only in the domain $0<x < +\infty$, since the solution at $-\infty < x < 0$ is given by means of the symmetry.    This is quite different from other wave equations such as the CH equation and the fully nonlinear wave equations, which are  defined in the {\em whole}  domain $-\infty < x < +\infty$.   Physically, it means that the flow at crest is {\em not} absolutely necessary to be irrotational.   Thus,  mathematically,  we need not consider whether the peaked/cusped solitary waves are weak solutions or not.  This is the reason why, unlike the well-known peaked solitary wave $\eta = c \; \exp(-|x|)$ of the CH equation, whose phase speed is {\em always} equal to wave height,  the phase speed of the peaked/cusped solitary waves (\ref{eta:peaked}) given by the UWM  has {\em nothing} to do with wave height!    This is the most attractive novelty of the UWM, which provides us a simple but elegant relationship between peaked and cusped solitary waves in finite water depth.         

This work is partly supported by the State Key Laboratory of Ocean Engineering (Approval No. GKZD010061) and the National Natural Science Foundation of China (Approval No. 11272209).


\bibliographystyle{unsrt}

\end{document}